\documentstyle{article}

\begin{document}

\begin{center}
{\large {\bf Local-Ether Wave Equation of Electric Field}}

{\large {\bf and Interferometry Experiments with Moving Medium and Path}}

$\ $

{\sf Ching-Chuan Su}

Department of Electrical Engineering

National Tsinghua University

Hsinchu, Taiwan
\end{center}

$\ $

\noindent {\bf Abstract} -- Recently, we have presented the local-ether
model, whereby the propagation of earthbound waves is supposed to be
referred uniquely to a geostationary inertial frame. Further, in order to
comply with this propagation model, the modified Lorentz force law is
developed. Thereby, the corresponding wave equations of potentials and
fields are derived in this investigation. It is shown that the local-ether
wave equation of electric field can account for various precision
interferometry experiments in a consistent way, including the one-way-link
experiment with a geostationary fiber, the Sagnac rotating-loop experiment
with a comoving or a geostationary dielectric medium, and Fizeau's
experiment with a moving dielectric medium in a geostationary
interferometer. These experiments together then provide a support for the
local-ether wave equation. Meanwhile, some other phenomena are predicted,
which provide a means to test its validity.

$\ $

$\ $

\noindent {\large {\bf 1. Introduction}}

Recently, we have presented the local-ether model of wave propagation
whereby electromagnetic wave is supposed to propagate via a medium like the
ether [1]. However, the ether is not universal. Specifically, it is proposed
that in the region under sufficient influence of the gravity due to the
Earth, the Sun, or another celestial body, there forms a local ether which
in turn moves with the gravitational potential of the respective body.
Thereby, each local ether together with the gravitational potential moves
with the associated celestial body. Thus, as well as earth's gravitational
potential, the earth local ether for earthbound propagation is stationary in
an ECI (earth-centered inertial) frame. Consequently, earthbound wave
phenomena can depend on earth's rotation but are entirely independent of
earth's orbital motion around the Sun or whatever. Meanwhile, the sun local
ether for interplanetary propagation is stationary in a heliocentric
inertial frame. This local-ether model has been adopted to account for the
effects of earth's motions in a wide variety of propagation phenomena,
particularly the Sagnac correction in GPS (global positioning system), the
Sagnac effect in rotating-loop interferometers, the time comparison via
intercontinental microwave link, and the echo time in interplanetary radar.
As examined within the present accuracy, the local-ether model is still in
accord with the Michelson-Morley experiment which is known to make the
classical ether notion obsolete. Moreover, by modifying the speed of light
in a gravitational potential, this simple propagation model leads to the
deflection of light by the Sun and the increment of echo time in the
interplanetary radar which are important phenomena supporting the general
theory of relativity [1].

Further, it is noticed that the electric current generating the magnetic
field is commonly electrically neutralized. That is, the mobile charged
particles which form the current are actually drifting in a matrix and the
ions which constitute the matrix tend to electrically neutralize the mobile
particles, such as electrons in a metal wire. Thus the velocity which
determines the current density is the drift velocity of the mobile particles
with respect to the neutralizing matrix. Consequently, the neutralized
current density remains unchanged when observed in different reference
frames. In order to comply with the local-ether propagation model and the
frame-independence of the current density, we have developed an
electromagnetic force law which complies with Galilean transformations while
it can reduce to a familiar form under some common condition [2]. In this
new classical theory all of the involved position vectors, time derivatives,
and velocities are referred specifically to their respective reference
frames. Owing to this simple feature, the associated potentials and
electromagnetic force remain unchanged when observed in different frames.
Under the common low-speed condition where the source particles forming the
current drift very slowly with respect to the neutralizing matrix, this
force law reduces to a form similar to the Lorentz force law. However, the
fundamental modification is that the current density generating the magnetic
field, the time derivative applied to the magnetic vector potential in the
electric induction force, and the particle velocity connecting to the
magnetic field in the magnetic force are all referred specifically to the
matrix frame in which the matrix is stationary. The modified equation is
identical to the Lorentz force law, if the latter is observed in the matrix
frame as done tacitly in common practice.

The propagation of the electromagnetic potentials is supposed to follow the
local-ether model. Accordingly, the wave equations of potentials are derived
in this investigation. Further, based on the modified Lorentz force law and
the associated definitions of electric and magnetic fields in terms of the
potentials, the local-ether wave equations of fields are derived. Thereby,
the phase speed and propagation constant of the electromagnetic wave
propagating in a moving medium are explored. Then, after the Sagnac effect
due to the movement of the propagation path is taken into consideration, we
present consistent reinterpretations for some precision interferometry
experiments, including the fiber-link experiment with a geostationary setup
and a geostationary optical fiber, the Sagnac rotating-loop experiments with
a comoving or a geostationary dielectric medium, and Fizeau's experiment
with a geostationary interferometer and a moving dielectric medium.

$\ $

\noindent {\large {\bf 2. Modified Lorentz Force Law}}

Based on Galilean transformations, we have presented an electromagnetic
force law expressed in terms of the augmented scalar potential, which is the
electric scalar potential enhanced slightly with a term associated with the
difference between the velocities of the effector and the source particles
[2]. Consider the electromagnetic force exerted on an effector particle of
charge $q$ due to an ensemble of source particles of charge density $\rho
_{v}$ drifting in a matrix of charge density $\rho _{m}$. The drift velocity
of the mobile source particles with respect to the matrix is given by the
difference ${\bf v}_{sm}={\bf v}_{s}-{\bf v}_{m}$, where ${\bf v}_{s}$ and $%
{\bf v}_{m}$ are the velocities of the source particles and the matrix,
respectively, both with respect to a particular frame. The matrix can be a
metal wire, a dielectric medium, or a magnet and tends to electrically
neutralize the mobile source particles to some extent.

Under the common low-speed condition where the speeds of the involved
particles are low and the source particles drift very slowly with respect to
the matrix, the electromagnetic force law proposed in terms of the augmented
potentials has been shown to reduce to the modified Lorentz force law [2] 
$$
{\bf F}({\bf r},t)=q\left\{ {\bf E}({\bf r},t)+{\bf v}_{em}\times {\bf B}(%
{\bf r},t)\right\} ,\eqno
(1) 
$$
where the velocity difference ${\bf v}_{em}$ ($={\bf v}_{e}-{\bf v}_{m}$) is
the velocity of the effector particle with respect to the matrix and ${\bf v}%
_{e}$ is the effector velocity with respect to the aforementioned particular
frame. The electric and magnetic fields in turn are defined explicitly in
terms of the potentials $\Phi $ and ${\bf A}$ as 
$$
{\bf E}({\bf r},t)=-\nabla \Phi ({\bf r},t)-\left( \frac{\partial }{\partial
t}{\bf A}({\bf r},t)\right) _{m}\eqno
(2) 
$$
and 
$$
{\bf B}({\bf r},t)=\nabla \times {\bf A}({\bf r},t),\eqno
(3) 
$$
where the time derivative $(\partial /\partial t)_{m}$ is referred
specifically to the matrix frame. Quantitatively, the electric scalar
potential $\Phi $ and the magnetic vector potential ${\bf A}$ are defined
explicitly as 
$$
\Phi ({\bf r},t)=\frac{1}{\epsilon _{0}}\int \frac{\rho _{n}({\bf r}^{\prime
},t-R/c)}{4\pi R}dv^{\prime }\eqno
(4) 
$$
and 
$$
{\bf A}({\bf r},t)=\frac{1}{\epsilon _{0}c^{2}}\int \frac{{\bf J}_{n}({\bf r}%
^{\prime },t-R/c)}{4\pi R}dv^{\prime },\eqno
(5) 
$$
where the net charge density $\rho _{n}=\rho _{v}+\rho _{m}$, the
neutralized current density ${\bf J}_{n}={\bf v}_{sm}\rho _{v}$, the
propagation range $R=|{\bf r}-{\bf r}^{\prime }|$, and the position vectors $%
{\bf r}$ and ${\bf r}^{\prime }$ are referred to a particular frame.

Obviously, the values of the neutralized current density and the potentials
remain unchanged when observed in different frames. Moreover, since the time
derivative in the definition of field ${\bf E}$ and the effector velocity in
the force law are referred uniquely to the matrix frame, the values of
fields ${\bf E}$ and ${\bf B}$ and hence the electromagnetic force exerted
on a charged particle given in the preceding formulas also remain unchanged
in different frames. That is, based on Galilean transformations, the values
of potentials, fields, and electromagnetic force are independent of
reference frame. Further, it is noted that the drift velocity ${\bf v}_{sm}$
involved in the neutralized current density is a Newtonian relative velocity
and hence this current density complies with Galilean relativity. For
quasi-static case where the propagation time $R/c$ in the potentials can be
neglected, the potentials $\Phi $ and ${\bf A}$ then comply also with
Galilean relativity and comove with the matrix.

It is seen that the force law (1) looks like the Lorentz force law. However,
the fundamental modification is that the current density generating the
magnetic vector potential, the time derivative applied to this potential in
the electric induction force, and the particle velocity connecting to the
curl of this potential in the magnetic force are all referred specifically
to the matrix frame. Meanwhile, in applying the Lorentz force law to the
analysis of, say, a motor or the magnetic deflection, the adopted reference
frame is usually the one with respect to which the magnetic field and the
magnet are stationary. Thus the matrix frame has been adopted tacitly as the
reference frame. Therefore, the modified equation is identical to the
Lorentz force law, if the latter is observed in the matrix frame as done
tacitly in common practice.

$\ $

\noindent {\large {\bf 3. Local-Ether Wave Equations of Potentials and Fields%
}}

Based on the local-ether model of wave propagation and the electromagnetic
force in terms of the potentials, it is supposed here that each individual
source particle continuously excites the electric scalar potential and hence
other local-ether potentials. These potentials in turn propagate radially
outward from the source position at an isotropic speed $c$ with respect to
the associated local-ether frame, independent of the motions of source and
effector. That is, the position vectors ${\bf r}$ and ${\bf r}^{\prime }$
and hence the propagation range $R$ in the potentials $\Phi $ and ${\bf A}$
defined in (4) and (5) are referred specifically to the local-ether frame
and the ratio $R/c$ represents the propagation time from the source point $%
{\bf r}^{\prime }$ at the instant $t^{\prime }$ ($=t-R/c$) of wave emission
to the field point ${\bf r}$ at the instant $t$. To comply with the frame of
the position vectors, the velocities ${\bf v}_{e}$, ${\bf v}_{s}$, and ${\bf %
v}_{m}$ are also referred to the local-ether frame.

By applying the Laplacian operator to both sides of the integral formulas
for potentials (4) and (5) and then expanding the Laplacians of the
time-dependent charge and current densities divided by $R$ [3], it can be
shown that the wave equations of these local-ether potentials are given by 
$$
\nabla ^{2}\Phi ({\bf r},t)-\frac{1}{c^{2}}\frac{\partial ^{2}}{\partial
t^{2}}\Phi ({\bf r},t)=-\frac{1}{\epsilon _{0}}\rho _{n}({\bf r},t)\eqno
(6) 
$$
and 
$$
\nabla ^{2}{\bf A}({\bf r},t)-\frac{1}{c^{2}}\frac{\partial ^{2}}{\partial
t^{2}}{\bf A}({\bf r},t)=-\mu _{0}{\bf J}_{n}({\bf r},t),\eqno
(7) 
$$
where $\mu _{0}=1/\epsilon _{0}c^{2}$ and the position vector ${\bf r}$
together with the time derivative $\partial /\partial t$ is referred to the
local-ether frame, the reference frame of the wave propagation.

Then, by manipulating the wave equations of potentials according to the
definitions of fields, the wave equations of fields can be derived. By so
doing, we have the local-ether wave equation of magnetic field 
$$
\nabla ^{2}{\bf B}({\bf r},t)-\frac{1}{c^{2}}\frac{\partial ^{2}}{\partial
t^{2}}{\bf B}({\bf r},t)=-\mu _{0}\nabla \times {\bf J}_{n}({\bf r},t)\eqno
(8) 
$$
and the local-ether wave equation of electric field 
$$
\nabla ^{2}{\bf E}({\bf r},t)-\frac{1}{c^{2}}\frac{\partial ^{2}}{\partial
t^{2}}{\bf E}({\bf r},t)=\frac{1}{\epsilon _{0}}\nabla \rho _{n}({\bf r}%
,t)+\mu _{0}\left( \frac{\partial }{\partial t}{\bf J}_{n}({\bf r},t)\right)
_{m},\eqno
(9) 
$$
where, again, the position vector ${\bf r}$ together with the time
derivative $\partial /\partial t$ is referred to the local-ether frame. It
is noted that in the preceding equation the time derivative applied to the
current density is referred to the matrix frame and the velocity of the
mobile source particles forming this current density is also referred to
this frame. In free space having no sources, the local-ether wave equations
take a simpler form of 
$$
\nabla ^{2}\Psi ({\bf r},t)-\frac{1}{c^{2}}\frac{\partial ^{2}}{\partial
t^{2}}\Psi ({\bf r},t)=0,\eqno
(10) 
$$
where $\Psi $ denotes the scalar potential $\Phi $ or any Cartesian
component of the vector potential ${\bf A}$, electric field ${\bf E}$, or
magnetic field ${\bf B}$. It is seen that both electric and magnetic fields
as well as the potentials propagate at the speed $c$ with respect to the
local-ether frame. These local-ether wave equations made their debut in [4].

Consider the case of wave propagation in a uniform magnetic medium of
permeability $\mu $, where ${\bf J}_{n}=\nabla \times {\bf M}$ for the
magnetization current and the magnetization vector ${\bf M}=(1/\mu
_{0}-1/\mu ){\bf B}$. Thereby, the wave equation of magnetic field becomes 
$$
\frac{\mu _{0}}{\mu }\nabla ^{2}{\bf B}({\bf r},t)-\frac{1}{c^{2}}\frac{%
\partial ^{2}}{\partial t^{2}}{\bf B}({\bf r},t)=0,\eqno
(11) 
$$
where we have made use of the relation $\nabla {\bf \cdot B}=0$, which in
turn is a consequence of (3). It is seen that in a uniform magnetic medium
of $\mu $, electromagnetic wave propagates at the speed $c\sqrt{\mu _{0}/\mu 
}$ with respect to the local-ether frame. In the preceding derivation, no
information about the motion of the magnetic medium is involved. Thus the
derived propagation speed is independent of the motion of a uniform magnetic
medium.

Consider the analogous case of wave propagation in a dielectric medium of
permittivity $\epsilon $, where 
$$
{\bf J}_{n}({\bf r},t)=\left( \frac{\partial }{\partial t}{\bf P}({\bf r}%
,t)\right) _{m}\eqno
(12) 
$$
for the polarization current and the polarization vector ${\bf P}=(\epsilon
-\epsilon _{0}){\bf E}$. Note that the polarization current density is
associated with the time derivative of the polarization vector with respect
to the matrix frame, since the displacement of the polarization charge under
the influence of electric field is relative to the ions forming the matrix
and the drift velocity ${\bf v}_{sm}$ of the source particles forming a
neutralized current is also referred to the matrix frame. Another
consequence pertinent to this drift velocity is that the conservation of
charge leads to another relation between ${\bf J}_{n}$ and $\rho _{v}$ of 
$$
\nabla \cdot {\bf J}_{n}({\bf r},t)=-\left( \frac{\partial }{\partial t}\rho
_{v}({\bf r},t)\right) _{m}.\eqno
(13) 
$$
This relation is just the continuity equation, except that the time
derivative applied to the charge density is referred specifically to the
matrix frame. Further, this matrix-frame continuity equation can be given in
terms of the net charge density as 
$$
\nabla \cdot {\bf J}_{n}({\bf r},t)=-\left( \frac{\partial }{\partial t}\rho
_{n}({\bf r},t)\right) _{m}.\eqno
(14) 
$$
This is owing to the fact that the matrix charge density $\rho _{m}$ is
time-independent in the matrix frame. The preceding continuity equation
leads to a relation between the polarization charge density and the
polarization vector as $\rho _{n}=-\nabla \cdot {\bf P}$, where we have made
use of the initial condition that a uniform polarization ($\nabla \cdot {\bf %
P}=0$) corresponds to the complete neutralization ($\rho _{n}=0$). Thereby,
by using electric field to express the net charge density and the
neutralized current density induced in a dielectric medium, the wave
equation of electric field becomes 
$$
\nabla ^{2}{\bf E}({\bf r},t)-\frac{1}{c^{2}}\frac{\partial ^{2}{\bf E}({\bf %
r},t)}{\partial t^{2}}=-\frac{1}{\epsilon _{0}}\nabla \nabla \cdot
[(\epsilon -\epsilon _{0}){\bf E}({\bf r},t)]+\frac{\epsilon -\epsilon _{0}}{%
\epsilon _{0}}\frac{1}{c^{2}}\left( \frac{\partial ^{2}{\bf E}({\bf r},t)}{%
\partial t^{2}}\right) _{m}.\eqno
(15) 
$$
It is noted that this wave equation involves two time derivatives of
electric field referred to different frames.

Remark that based on Galilean transformations, the time derivatives of an
arbitrary function $f$ of space and time with respect to the matrix and the
local-ether frames are related by 
$$
\left( \frac{\partial f}{\partial t}\right) _{m}=\frac{\partial f}{\partial t%
}+{\bf v}_{m}\cdot \nabla f,\eqno
(16) 
$$
where the time derivatives $\partial /\partial t$ and $(\partial /\partial
t)_{m}$ are understood to be taken under constant ${\bf r}$ and (${\bf r}-%
{\bf v}_{m}t$), respectively, as the position vector ${\bf r}$ is referred
to the local-ether frame. Further, from the preceding relation, the
second-order time derivative in the matrix frame can be given in terms of
the derivatives in the local-ether frame as 
$$
\left( \frac{\partial ^{2}f}{\partial t^{2}}\right) _{m}=\frac{\partial ^{2}f%
}{\partial t^{2}}+2\left( {\bf v}_{m}\cdot \nabla \right) \frac{\partial f}{%
\partial t}+\left( {\bf v}_{m}\cdot \nabla \right) \left( {\bf v}_{m}\cdot
\nabla \right) f.\eqno
(17) 
$$
The Galilean formulas (16) and (17) have been given in [5] and [6],
respectively.

Then, by using the Galilean formula (17), the wave equation of electric
field can be rewritten as 
\[
\nabla ^{2}{\bf E}({\bf r},t)-\frac{\epsilon }{\epsilon _{0}}\frac{1}{c^{2}}%
\frac{\partial ^{2}{\bf E}({\bf r},t)}{\partial t^{2}}=-\frac{1}{\epsilon
_{0}}\nabla \nabla \cdot [(\epsilon -\epsilon _{0}){\bf E}({\bf r},t)]\ \ \
\ \ \ \ \ \ \ \ \ \ \ \ \ \ \ \ \ 
\]
$$
\ \ \ \ \ \ \ \ \ \ +\frac{\epsilon -\epsilon _{0}}{\epsilon _{0}}\frac{1}{%
c^{2}}\left[ 2\left( {\bf v}_{m}\cdot \nabla \right) \frac{\partial {\bf E}(%
{\bf r},t)}{\partial t}+\left( {\bf v}_{m}\cdot \nabla \right) \left( {\bf v}%
_{m}\cdot \nabla \right) {\bf E}({\bf r},t)\right] .\eqno
(18) 
$$
It is noted that this wave equation of electric field involves the matrix
velocity with respect to the local-ether frame. This feature is quite
different from that in the wave equation of magnetic field (11). Physically,
this difference is due to the situation that polarization current is
associated with the matrix-frame time derivative of field, while
magnetization current is with a space derivative. In terms of the
matrix-frame time derivatives, the local-ether wave equation of electric
field can be rewritten as 
$$
\nabla ^{2}{\bf E}-\frac{\epsilon }{\epsilon _{0}}\frac{1}{c^{2}}\left( 
\frac{\partial ^{2}{\bf E}}{\partial t^{2}}\right) _{m}=-\frac{1}{\epsilon
_{0}}\nabla \nabla \cdot [(\epsilon -\epsilon _{0}){\bf E}]-\frac{2}{c^{2}}%
\left( {\bf v}_{m}\cdot \nabla \right) \left( \frac{\partial {\bf E}}{%
\partial t}\right) _{m},\eqno
(19) 
$$
where the second-order term of the normalized speed $v_{m}/c$ is ignored. It
is seen that the last term in the preceding matrix-frame wave equation is
smaller in magnitude than the other terms by a factor of the order of the
normalized speed $v_{m}/c$.

Consider the simpler case where a $z$-polarized uniform plane wave
propagates along the $x$ direction in a uniform dielectric medium. The
medium is moving at a velocity ${\bf v}_{f}$ with respect to a laboratory
frame which in turn is moving at a velocity ${\bf v}_{0}$ with respect to
the local-ether frame. Based on Galilean transformations, the matrix
velocity with respect to the local-ether frame is simply ${\bf v}_{m}={\bf v}%
_{f}+{\bf v}_{0}$. Then, to the first order of normalized speed, the wave
equation (18) can be simplified to a form of 
$$
\frac{\partial ^{2}}{\partial x^{2}}E_{z}(x,t)-\frac{\epsilon }{\epsilon _{0}%
}\frac{1}{c^{2}}\frac{\partial ^{2}}{\partial t^{2}}E_{z}(x,t)=2v_{mx}\frac{%
\epsilon -\epsilon _{0}}{\epsilon _{0}}\frac{1}{c^{2}}\frac{\partial ^{2}}{%
\partial x\partial t}E_{z}(x,t),\eqno
(20) 
$$
where $v_{mx}={\bf v}_{m}\cdot \hat{x}$. It is noted that the term $\nabla
\cdot (\epsilon -\epsilon _{0}){\bf E}$ associated with the polarization
charge vanishes, since both permittivity $\epsilon $ and field ${\bf E}$ are
supposed to have no variations along the $z$ direction of the field.

Suppose that the solution of this wave equation is of a harmonic form as $%
E_{z}=E_{0}e^{ikx}e^{-i\omega t}$, where $x$ is referred to the local-ether
frame and $E_{0}$ is an arbitrary constant. Then the preceding wave equation
immediately leads to an algebraic equation 
$$
k^{2}-\frac{\epsilon }{\epsilon _{0}}\frac{\omega ^{2}}{c^{2}}=-2v_{mx}\frac{%
\epsilon -\epsilon _{0}}{\epsilon _{0}}\frac{1}{c^{2}}k\omega .\eqno
(21) 
$$
It is easy to show that to the first order of normalized speed, the
propagation constant $k$ can be given in terms of the angular frequency $%
\omega $ as 
$$
k=\frac{\omega }{c}\left\{ \sqrt{\frac{\epsilon }{\epsilon _{0}}}-\frac{%
v_{mx}}{c}\frac{\epsilon -\epsilon _{0}}{\epsilon _{0}}\right\} .\eqno
(22) 
$$
It is noted that this propagation constant depends on the speed $v_{mx}$, in
addition to the familiar dependence on permittivity $\epsilon $.
Consequently, this relation presents a modification of the propagation
constant due to the motion of the dielectric medium with respect to the
local ether. Physically, the dependence on the matrix speed originates from
the situation that the polarization current and its time derivative in the
wave equation are referred to the matrix frame rather than to the
local-ether one.

The phase speed $c^{\prime }$ ($=\omega /k$) of the electromagnetic wave
propagating in the moving dielectric medium is then given by 
$$
c^{\prime }=\frac{c}{n}+v_{mx}\left( 1-\frac{1}{n^{2}}\right) ,\eqno
(23)
$$
where the phase speed is referred to the local-ether frame and the
refractive index $n=\sqrt{\epsilon /\epsilon _{0}}$. It is seen that the
component of the matrix velocity parallel to the propagation direction
affects the phase speed, while the transverse components do not.

The wave equation (20) can also be written in the matrix frame. To the first
order of normalized speed, this wave equation reads 
$$
\frac{\partial ^{2}}{\partial x^{2}}E_{z}(x,t)-\frac{\epsilon }{\epsilon _{0}%
}\frac{1}{c^{2}}\left( \frac{\partial ^{2}}{\partial t^{2}}E_{z}(x,t)\right)
_{m}=-2v_{mx}\frac{1}{c^{2}}\left( \frac{\partial ^{2}}{\partial x\partial t}%
E_{z}(x,t)\right) _{m}.\eqno
(24) 
$$
It can be shown that $k=(\omega _{m}/c)(n+v_{mx}/c)$ and $c_{m}^{\prime
}=\omega _{m}/k=c/n-v_{mx}/n^{2}$, where $\omega _{m}$ and $c_{m}^{\prime }$
are the angular frequency and the phase speed of the wave when observed in
the matrix frame, respectively. It is seen that 
$$
c_{m}^{\prime }=c^{\prime }-v_{mx},\eqno
(25) 
$$
which states that the difference in the observed phase speed between the two
reference frames is just the relative speed between the frames along the
propagation direction. In other words, the derived phase speeds in the two
frames comply with Galilean transformations.

The phase speed given in (23) is similar to the speed obtained from the
velocity transformation in the special relativity, which in turn is known to
have been demonstrated in the famous Fizeau's interferometry experiment with
flowing water. However, the fundamental difference is that the phase speed $%
c^{\prime }$ and the matrix velocity ${\bf v}_{m}$ are referred specifically
to the local-ether frame. Thus the matrix velocity ${\bf v}_{m}$
incorporates the laboratory velocity ${\bf v}_{0}$ with respect to the
local-ether frame. An old interpretation of the dependence of the phase
speed on the medium velocity given in (23) is known as the Fresnel drag
effect [6, 7], whereby the ether is dragged partially by the moving medium
to an extent depending on the index $n$. If $n$ is large enough, $c^{\prime
}\simeq c/n+v_{mx}$ and $c_{m}^{\prime }\simeq c/n$. It is seen that the
phase speed with respect to the moving medium is independent of the velocity
of the medium. Thus it seems that the ether is completely dragged by the
medium. However, according to the wave equation of electric field (15), the
earth local ether is still stationary in an ECI frame, regardless of the
permittivity and the velocity of a terrestrial medium. The change in the
phase speed of a wave propagating in a dielectric medium, stationary or
moving, is due to the effect of the polarization which in turn is induced by
and related to electric field. Further, it is noted that the interpretation
similar to the ether dragging is applicable only for a uniform plane wave
propagating in a uniform dielectric medium and is correct only to the first
order. In other words, according to the local-ether wave equation, the
phase-speed formula (23) does not hold in a magnetic medium and can be no
longer valid in a nonuniform dielectric medium where polarization charge
emerges. Thus the applicability of the phase-speed formula has some hidden
restrictions. This presents a viewpoint different from the ether-dragging
notion and another discrepancy from the special relativity.

$\ $

\noindent {\large {\bf 4. Reexamination of Various Interferometry Experiments%
}}

In this section we reexamine some precision interferometry experiments
reported in the literature to test the local-ether wave equation of electric
field. They include the fiber-link experiment with a geostationary setup and
a geostationary optical fiber, the Sagnac loop interferometry with a
rotating path and a comoving or geostationary dielectric medium, and
Fizeau's experiment with a geostationary interferometer and a moving
dielectric medium. For these earthbound experiments, the local ether is
stationary in an ECI frame and hence earth's rotation should be taken into
consideration, even for a geostationary medium or path. On the other hand,
these experiments are entirely independent of earth's orbital motion around
the Sun or others.

$\ $

\noindent {\bf 4.1. Phase variation with moving medium and path}

First of all, consider an interferometer of which each segment of
propagation path has a fixed shape and is implemented with a pair of mirrors
in air, with a pipe filled with flowing dielectric liquid, or with a solid
dielectric fiber. Remark that for an electromagnetic wave propagating in an
arbitrary direction $\hat{l}$ along a path filled with a dielectric medium
moving at a velocity ${\bf v}_{m}$ with respect to the local-ether frame,
the propagation constant can be written as 
$$
k=k_{0}\left\{ n+\left( 1-n^{2}\right) \hat{l}\cdot {\bf v}_{m}/c\right\} ,%
\eqno
(26) 
$$
where $k_{0}=\omega /c$ and $n$ is the refractive index of the uniform
moving medium.

Further, consider a propagation path segment of length $l$, such as a linear
section of the pipe or fiber. Owing to the movement of the segment during
the wave propagation through it, the propagation range which represents the
actual propagation length is not actually $l$. As in the classical
propagation model, the propagation range $l^{\prime }$ depends on the
velocity of the path segment with respect to the local ether. The influence
on propagation due to the difference between the propagation range $%
l^{\prime }$ and the path length $l$ is known as the Sagnac effect. By
following the derivation of the Sagnac effect for electromagnetic wave
propagating in free space discussed elaborately in [1], the propagation
range $l^{\prime }$ for a wave propagating in the direction $\hat{l}$ along
a moving segment, when given to the first order of normalized speed, is 
$$
l^{\prime }=l\left\{ 1+n\hat{l}\cdot {\bf v}_{l}/c\right\} ,\eqno
(27) 
$$
where ${\bf v}_{l}$ is the velocity of the path segment with respect to the
local-ether frame.

It is supposed that the phase variation over the propagation path is given
by $\phi =kl^{\prime }$. To the first order of normalized speed, the phase
variation associated with the moving medium and path is then given from the
two preceding formulas by 
$$
\phi =k_{0}l\left\{ n+(1-n^{2})\hat{l}\cdot {\bf v}_{m}/c+n^{2}\hat{l}\cdot 
{\bf v}_{l}/c\right\} .\eqno
(28) 
$$
It is seen that the phase variation depends on the longitudinal (to the
propagation path) $\hat{l}$ components of the medium velocity and the path
velocity both with respect to the local-ether frame. Furthermore, these
longitudinal-speed terms connect with the refractive index. Thereby, in
addition to the familiar effect on the propagation constant, the refractive
index also has effects on the matrix-velocity modification of the
propagation constant and the path-velocity modification of the propagation
length. The preceding phase-variation formula can be rewritten as 
$$
\phi =k_{0}l\left\{ n+n^{2}\hat{l}\cdot {\bf v}_{lm}/c+\hat{l}\cdot {\bf v}%
_{m}/c\right\} ,\eqno
(29) 
$$
where ${\bf v}_{lm}={\bf v}_{l}-{\bf v}_{m}$ denotes the Newtonian relative
velocity of the moving path with respect to the moving dielectric medium. It
is seen that one of the velocity-dependent terms depends on the refractive
index and is associated with a Newtonian relative velocity, while the other
velocity-dependent term is independent of the index and is associated with
the individual velocity of the dielectric medium. For the case where the
medium comoves with the propagation path (${\bf v}_{lm}=0$), the
velocity-dependent phase variation is no longer dependent on the index.

$\ $

\noindent {\bf 4.2. Geostationary fiber-link experiment}

Then we consider the experiment of one-way fiber link by Krisher {\it et al}%
., where the phase difference between two waves generated from two identical
stable hydrogen masers at 100 MHz was measured by using a network analyzer
[8]. The two masers are separated by a long distance of 21 km and linked by
a stable optical fiber. One signal was fed directly into the analyzer, while
the other was used to modulate a laser carrier to propagate over the fiber.
For a geostationary fiber, ${\bf v}_{m}={\bf v}_{l}={\bf v}_{E}$, where $%
{\bf v}_{E}$ is the linear velocity due to earth's rotation. Then formula
(29) immediately leads to that the phase variation $d\phi $ over a
geostationary fiber segment of differential length $dl$ is given by 
$$
d\phi =k_{0}\left( n+\hat{l}\cdot {\bf v}_{E}/c\right) dl.\eqno
(30) 
$$
Thus the phase variation over a propagation path $L$ of length $l$ is given
by the path integral 
$$
\phi =k_{0}\left( nl+\frac{1}{c}\int_{L}{\bf v}_{E}\cdot d{\bf l}\right) 
{\bf ,}\eqno
(31) 
$$
where the directed differential length $d{\bf l}=\hat{l}dl$. It is noted
that the velocity-dependent part is independent of the refractive index,
since the velocity difference ${\bf v}_{lm}=0$. Furthermore, it is noted
that although the velocity-dependent phase variation is due to earth's
rotation, its value is invariant under this rotation. This is because that
both ${\bf v}_{E}$ and $d{\bf l}$ change in a coordinated way such that
their dot product remains unchanged during earth's rotation. This invariance
is in accord with the spatial isotropy demonstrated in the one-way
fiber-link experiment, where it is observed that the phase variation is
highly stable (as determined from the phase difference between the two waves
measured every a few seconds during a couple of days), regardless of earth's
rotational and orbital motions [8].

When the fiber link is placed in a geographically small region such that the
velocity ${\bf v}_{E}$ due to earth's rotation is substantially constant,
the phase variation over the link becomes 
$$
\phi =k_{0}\left( nl+\frac{1}{c}{\bf v}_{E}\cdot {\bf d}\right) {\bf ,}%
\eqno
(32)
$$
where ${\bf d}$ is the directed separation distance from the end through
which the light enters the fiber to the other end. It is noted that the
phase variation is independent of the actual wiring, but depends on the
orientation of the distance ${\bf d}$ with respect to the ground. If the
fiber-link experimental setup is put on a turntable or a rotor, it is
expected that the phase variation will change as the turntable is rotating
even slowly. From the minor term in the preceding formula, the dependence of
the phase variation on the orientation of the rotor is given by 
$$
\delta \phi =\frac{\omega }{c^{2}}\omega _{E}R_{E}|{\bf d|}\cos \theta
_{l}\cos \theta _{T}{\bf ,}\eqno
(33)
$$
where $R_{E}$ is earth's radius, $\theta _{l}$ is the latitude, $\theta _{T}$
is the angle of the distance ${\bf d}$ from the east, and the turntable is
suitably positioned on a horizontal plane so that the gravitational effect
on $\omega $ is avoided. The phase variation associated with earth's
rotation is independent of the refractive index of the fiber and hence is
identical to the one in the case with a free-space link discussed in [1]. It
is noted that the phase variation becomes maximum and minimum, as the
distance ${\bf d}$ points to the east and the west, respectively.
Furthermore, the phase variation changes sinusoidally with the angle $\theta
_{T}$. This dipole anisotropy in phase variation predicted in the proposed
one-way-link rotor experiment then provides a means to test the local-ether
wave equation.

$\ $

\noindent {\bf 4.3. Sagnac rotating-loop experiments}

Consider the Sagnac effect in a rotating-loop interferometer associated with
the interference between two coherent waves corotating and counterrotating
with respect to the propagation loop, respectively. Commonly, the
propagation path is composed of beam splitter and mirrors in air. Then the
phase difference between the two waves results from the situation that the
modification of propagation length due to the rotation of the loop is
different between these waves propagating in opposite directions. Here, we
deal with a propagation path filled with a general dielectric medium rather
than a free space.

Suppose that the loop is rotating about an axis as observed from a
laboratory, which in turn is rotating with the Earth. The velocity of the
path segment with respect to the an ECI frame is given by ${\bf v}_{l}={\bf v%
}_{I}+{\bf v}_{E0}+{\bf v}_{0}$, where ${\bf v}_{0}$ is the laboratory
velocity with respect to an ECI frame referred to a suitable point ${\bf r}%
_{0}$ on the interferometer rotation axis, ${\bf v}_{I}=\bar{\omega}%
_{I}\times ({\bf r}-{\bf r}_{0})$ and ${\bf v}_{E0}=\bar{\omega}_{E}\times (%
{\bf r}-{\bf r}_{0})$ denote the other contributions to the path velocity
due to the rotations of the loop and the Earth at the directed rotation
rates $\bar{\omega}_{I}$ and $\bar{\omega}_{E}$, respectively, and ${\bf r}$
is the position vector of the path segment in the frame of the reference
point ${\bf r}_{0}$.

Consider the case where the medium is comoving with the rotating loop. Thus
the velocity difference ${\bf v}_{lm}=0$ and hence formula (29) leads to
that the phase variation $d\phi $ over a propagation path of differential
length $dl$ is given by 
$$
d\phi =k_{0}\left( n+\hat{l}\cdot {\bf v}_{l}/c\right) dl.\eqno
(34) 
$$
The unit vectors $\hat{l}$ denoting the propagation directions of the two
counterpropagating waves are antiparallel to each other in each path
segment. Thus the difference in the phase variation over the loop $L$
between the two waves is then given by the path integral 
$$
\triangle \phi =\frac{2k_{0}}{c}\oint_{L}({\bf v}_{I}+{\bf v}_{E0})\cdot d%
{\bf l,}\eqno
(35) 
$$
where we have made use of the facts that the major term $k_{0}ndl$ in (34)
is identical for the two waves and hence its contributions to the phase
difference cancel out and that the contributions of a constant vector ${\bf v%
}_{0}$ to the phase variation over a closed path cancel out collectively,
regardless of the actual structure of the loop.

Suppose the loop is coplanar. Then, by using a vector identity, the phase
difference can be given by 
$$
\triangle \phi =\frac{4k_{0}}{c}(\bar{\omega}_{I}+\bar{\omega}_{E})\cdot 
{\bf S,}\eqno
(36) 
$$
where ${\bf S}$ ($=\frac{1}{2}\oint_{L}({\bf r}-{\bf r}_{0})\times d{\bf l}$%
) denotes the directed area enclosed by loop $L$. It is noted that the phase
difference is independent of the index $n$ and hence is identical to the one
in the case with a free-space path discussed in [1]. Again, this
independence is owing to the coincidence that the index-dependent
modification of the propagation constant happens to cancel the one of the
propagation length. This null effect of the refractive index on the phase
difference in the Sagnac loop interferometry has been demonstrated
experimentally by Harzer in as early as 1914 [9]. Further, the preceding
phase-difference formula has been put in practical use in fiber gyroscopes
[10-12]. Alternative derivations of this formula by taking the Sagnac effect
and the modifications of phase speed into account can be found in [9-11].
Anyway, in spite of the restriction on reference frame, the local-ether wave
equation of electric field is in accord with the Sagnac interferometry
experiment with a rotating loop and a comoving dielectric medium. On the
other hand, the local-ether wave equation precludes the possibility of
detecting earth's orbital motion by using earthbound fiber gyroscopes.

Next, consider the case where the medium is geostationary while the
propagation loop is still rotating with respect to the laboratory. Thus $%
{\bf v}_{lm}={\bf v}_{I}$, ${\bf v}_{m}={\bf v}_{E0}+{\bf v}_{0}$, and hence
formula (29) leads to that the phase variation $d\phi $ over a path of
differential length $dl$ is given by 
$$
d\phi =k_{0}dl\left( n+n^{2}\hat{l}\cdot {\bf v}_{I}/c+\hat{l}\cdot {\bf v}%
_{E0}/c+\hat{l}\cdot {\bf v}_{0}/c\right) .\eqno
(37)
$$
The corresponding phase difference between the waves becomes 
$$
\triangle \phi =\frac{4k_{0}}{c}(n^{2}\bar{\omega}_{I}+\bar{\omega}%
_{E})\cdot {\bf S}.\eqno
(38)
$$
It is seen that the phase difference then depends on the index $n$ which in
turn connects to the loop rotation rate $\bar{\omega}_{I}$. Again, the phase
variation associated with earth's rotation is independent of the index and
the phase difference is independent of the laboratory velocity ${\bf v}_{0}$%
. Ordinarily, $\omega _{E}\ll \omega _{I}$ and hence the phase difference is
substantially proportional to $n^{2}$. This index-dependence of the phase
difference is identical to that given in [9], although the approach is quite
different. The increase of phase difference in the presence of a stationary
dielectric medium has been demonstrated experimentally by Dufour and Prunier
in 1942 [9]. The various index-dependences among the terms connected to the
rate $\bar{\omega}_{I}$ or $\bar{\omega}_{E}$ in (36) and (38) then provide
another means to test the local-ether wave equation.

$\ $

\noindent {\bf 4.4. Fizeau's experiment with moving medium}

The last case is then the one where the interferometer is stationary while
the medium is moving. Consider Fizeau's experiment dealing with the
interference between two optical waves propagating in opposite directions
along a stationary pipe filled with flowing water. For generality, suppose
that the pipe together with beam splitters, mirrors, and other components of
the interferometer is stationary in a laboratory frame which in turn moves
at a velocity ${\bf v}_{0}$ with respect to an ECI frame. As the
interferometer is geographically small, the path velocity is substantially
constant over the setup. Thus the path velocity ${\bf v}_{l}={\bf v}_{0}$.
Further, suppose that the water is flowing at a velocity ${\bf v}_{f}$ with
respect to the pipe. Then the velocity difference ${\bf v}_{lm}=-{\bf v}_{f}$%
, the matrix velocity ${\bf v}_{m}={\bf v}_{f}+{\bf v}_{0}$, and hence
formula (29) leads to that the phase variation $d\phi $ over a pipe of
differential length $dl$ is given by 
$$
d\phi =k_{0}dl\left\{ n+(1-n^{2})\hat{l}\cdot {\bf v}_{f}/c+\hat{l}\cdot 
{\bf v}_{0}/c\right\} ,\eqno
(39) 
$$
where $n$ is the index of the flowing water and $\hat{l}\cdot {\bf v}%
_{f}=\pm v_{f}$ for the two beams.

It is seen that the phase variation depends on the laboratory velocity.
However, the effect of this velocity on phase variation can not be detected
in Fizeau's experiment. This is because that the optical path actually
adopted in the experiment is closed, part of the path is filled with flowing
water and part is merely with air (see the figure in [13] or [14]). It is
noted that in the preceding formula the velocity ${\bf v}_{0}$ does not
connect to the index $n$. Again, the phase variation over a closed path
given by the circulation of a constant vector ${\bf v}_{0}$ is zero.

As in the Sagnac loop interferometry, the two waves to be interfered
propagate in opposite directions in each individual segment of the path.
From the preceding formula it is seen that those parts of the path filled
with air or with a dielectric at rest with the pipe do not contribute to the
phase difference between the waves. The contribution comes only from the
water-flowing pipe. Suppose that the water speed $v_{f}$ is uniform and the
total length of the water-flowing pipe is $l$. Then the difference in phase
variation between the two counterpropagating waves is given by 
$$
\triangle \phi =2k_{0}l(n^{2}-1)v_{f}/c.\eqno
(40) 
$$
It is noted that the phase difference is linearly proportional to ($n^{2}-1$%
) and to the water speed $v_{f}$ with respect to the pipe. The preceding
phase-difference formula agrees with the interference fringe observed in
Fizeau's experiment with various speeds $v_{f}$ and indices $n$.

More precisely, the path velocity ${\bf v}_{l}$ is not actually a constant
value ${\bf v}_{0}$ over the closed path. That is, ${\bf v}_{l}={\bf v}_{E0}+%
{\bf v}_{0}$ and ${\bf v}_{m}={\bf v}_{f}+{\bf v}_{E0}+{\bf v}_{0}$, where $%
{\bf v}_{0}$ is referred to a suitable point in the setup. Hence the Sagnac
effect in a loop interferometer due to earth's rotation should also appear
in Fizeau's experiment. However, as earth's rotation rate is relatively
slow, its effect in Fizeau's experiment is ordinarily much smaller than that
due to the motion of the medium, just as its effect in the rotating-loop
experiment is ordinarily much smaller than that due to the rotation of the
loop. Thereby, the phase difference is substantially independent of earth's
rotation. Furthermore, at least to the first order of normalized speed,
Fizeau's experiment together with the Sagnac rotating-loop experiments\ is
independent of the laboratory velocity ${\bf v}_{0}$ and hence complies with
Galilean relativity, in spite of the restriction on reference frame of the
medium and the path velocities.

$\ $

\noindent {\large {\bf 5. Conclusion}}

Based on the local-ether model of wave propagation, the propagation of the
potentials is referred specifically to an ECI frame in earthbound
experiments. Further, under the ordinary condition of low drift speed, the
electromagnetic force can be given in terms of electric and magnetic fields
which in turn are given explicitly in terms of the local-ether potentials.
The position vectors, time derivatives, particle velocities, propagation
velocity, and current density involved are all referred specifically to
their respective reference frames. Consequently, the values of potentials,
fields, and force remain unchanged in different frames.

Further, based on the definitions of fields in terms of potentials, the
local-ether wave equations of fields are derived. From the wave equation of
magnetic field, the phase speed of electromagnetic wave in a uniform
magnetic medium is found to depend on the permeability, but not on the
motion of the medium. However, from the wave equation of electric field, the
phase speed of a uniform plane wave propagating in a moving uniform
dielectric medium is found to depend on the longitudinal component of the
medium velocity and hence to incorporate the familiar Fresnel drag
coefficient. This phase speed looks like the speed obtained from the
velocity transformation in the special relativity. However, the fundamental
difference is that the phase and the matrix speeds are referred specifically
to the local-ether frame. Moreover, this phase-speed formula is not expected
to hold in a magnetic or a nonuniform dielectric medium.

By taking the matrix-velocity modification of the propagation constant and
the Sagnac path-velocity modification of the propagation length into
account, the phase variation over a moving path filled with a moving
dielectric medium is presented. Thereby, this phase-variation formula is
applied to analyze various precision interferometry experiments in a
consistent way. It is found that this formula is actually in accord with the
spatial isotropy in the one-way fiber-link experiment, with the null effect
of permittivity in the Sagnac interferometer with a dielectric medium
comoving with the rotating loop, with the increase of phase difference in
the Sagnac interferometer with a geostationary dielectric medium, and with
the dependence of phase difference on the speed and index of the flowing
water in Fizeau's experiment. These together provide a support for the
local-ether wave equation of electric field. Moreover, it is predicted that
as the fiber-link experimental setup is put on a turntable, the phase
variation over the link will change sinusoidally as the orientation of
turntable is changing. This proposed one-way-link rotor experiment, the
predicted null effect of earth's orbital motion in earthbound
interferometers, the predicted various index-dependences among the
phase-shift terms connected to the rotation rate of the loop or of the Earth
in the Sagnac interferometer, and the aforementioned discrepancies in the
phase-speed formula for a moving medium and the restrictions on this formula
then provide different approaches to test the local-ether wave equation.

$\ $

$\ $

\noindent {\large {\bf References}}

\begin{itemize}
\item[{\lbrack 1]}]  C.C. Su, {\it Eur. Phys. J. C} {\bf 21},{\rm \ }701
(2001); {\it Europhys. Lett}. {\bf 56}, 170 (2001).

\item[{\lbrack 2]}]  C.C. Su, in {{\it IEEE Antennas Propagat. Soc. Int}. 
{\it Symp}. {\it Dig.}} (2001), vol. 1, p. 208.

\item[{\lbrack 3]}]  C.C. Su, {\it J. Electromagnetic Waves Applicat.} {\bf %
14}, 1251 (2000).

\item[{\lbrack 4]}]  C.C. Su, in {\it IEEE Antennas Propagat. Soc. Int.
Symp. Dig}. (2001), vol. 1, p. 212.

\item[{\lbrack 5]}]  L.E. Ballentine, {\it Quantum Mechanics}
(Prentice-Hall, Englewood Cliffs, 1990), sect. 4-3.

\item[{\lbrack 6]}]  {J.D. Jackson, {\it Classical Electrodynamics} }({%
Wiley, New York, 1975), ch. 11.}

\item[{\lbrack 7]}]  See, for example, W.M. Macek, J.R. Schneider, and R.M.
Salamon, {\it J. Appl. Phys}. {\bf 35}, 2556 (1964).

\item[{\lbrack 8]}]  {T.P. Krisher, L. Maleki, G.F. Lutes, L.E. Primas, R.T.
Logan, J.D. Anderson, and C.M. Will, {\it Phys. Rev. D} {\bf 42}, 731 (1990).%
}

\item[{\lbrack 9]}]  E.J. Post, {\it Rev. Modern Phys}. {\bf 39}, 475 (1967).

\item[{\lbrack 10]}]  H.J. Arditty and H.C. Lef\`{e}vre, {\it Opt. Lett}. 
{\bf 6}, 461 (1981).

\item[{\lbrack 11]}]  S. Ezekiel, S.P. Smith, and F. Zarinetchi, in {\it %
Optical Fiber Rotation Sensing}, W.K. Burns ed. (Academic Press, New York,
1994), p. 4.

\item[{\lbrack 12]}]  B. Culshaw and I.P. Giles, {\it J. Phys. E} {\bf 16},
5 (1983).

\item[{\lbrack 13]}]  A.P. French{, {\it Special Relativity}} ({Chapman \&
Hall, New York, 1968), ch. 2.}

\item[{\lbrack 14]}]  In {\it McGraw-Hill Encyclopedia of Science }\&{\it \
Technology} (McGraw-Hill, {New York, }1992), vol. 10, p. 55 (in the article
of ``Light'').
\end{itemize}

\end{document}